\documentclass[preprint,12pt,3p]{elsarticle}

\usepackage{amssymb}
\usepackage{amsmath}

\begin{document}

\begin{frontmatter}

\title{Lagrangian Formalism in Biology: I. Standard Lagrangians
and their Role in Population Dynamics}

\author{D.T. Pham$^1$ and Z.E. Musielak$^2$}
\address{$^1$Department of Biology, University of Texas 
at Arlington, Arlington, TX 76019, USA \\}
\address{$^2$Department of Physics, University of Texas 
at Arlington, Arlington, TX 76019, USA \\}

\begin{abstract}
The Lagrangian formalism is developed for the population 
dynamics of interacting species that are described by several
well-known models.  The formalism is based on standard 
Lagrangians, which represent differences between the 
physical kinetic and potential energy-like terms.  A method 
to derive these Lagrangians is presented and applied to 
selected theoretical models of the population dynamics.
The role of the derived Lagrangians and the energy-like 
terms in the population dynamics is investigated and 
discussed.  It is suggested that the obtained standard 
Lagrangians can be used to identify physical similarities 
between different population models.
\end{abstract}


\end{frontmatter}

\section{Introduction}

The astounding progress of modern physics in discovering and 
understanding the fundamental laws of Nature that govern the 
structure and evolution of nonorganic matter has been caused 
by its powerful mathematical-empirical approach.  In this 
approach, physical theories are formulated using the language
of mathematics and their predictions are verified by experiments.
At the same time, life sciences, included biology, remain mainly 
empirical and descriptive.  

There have been attempts to formulate mathematical models of 
some biological systems, and thereby to establish mathematically 
oriented theoretical biology [1].  Different areas of mathematics 
have become increasingly more important in biology in recent 
decades, specifically, statistics in experimental design, pattern 
recognition in bioinformatics, and mathematical modeling in 
evolution, ecology and epidemiology [2]; however, it is pointed 
out that some of these attempts can be classified as 'uses' but 
others must be considered as 'abuses'.  

By combining theoretical and empirical biology, the
mathematical-empirical approach may be established with 
expectations that its description of living organisms reaches 
the level of the description given by physics to nonorganic matter.  
In modern theoretical physics, all fundamental equations describing 
matter are derived by using the Lagrangian formalism [3-5], which 
requires a prior knowledge of functions called Lagrangians [6-8].  
A number of different methods have been proposed [9-14] to 
obtain the Lagrangians for the most basic equations of modern 
physics [5].  

In theoretical biology, Kerner [15] was first who applied the 
Lagrangian formalism to biology and obtained Lagrangians 
for several selected biological systems described by first-order 
ordinary differential equations (ODEs).  Later, Paine [16] 
investigated the existence and construction of Lagrangians 
for similar set of ODEs following the original work of 
Helmholtz [6].  More recently, Nucci and Tamizhmani [17] 
derived Lagrangians for some models representing the 
population dynamics by using the method based on Jacobi 
Last Multiplier [10]. Moreover, Nucci and Sanchini [18] 
obtained Lagrangian for an Easter Island population model.   

The main goal of this paper is to develop the Lagrangian 
formalism for biological systems.  The formalism is based 
on standard Lagrangians, whose main characteristic is the 
presence of the difference between the kinetic and potential 
energy-like terms [5,9,10].  A method to derive these 
Lagrangians is presented and applied to selected models of
the population dynamics.  The role of the derived Lagrangians 
and the energy-like terms in these models is investigated and 
discussed.  It is suggested that the obtained standard Lagrangians 
can be used to identify physical similarities between diverse 
biological systems.

The paper is organized as follows. Section 2 presents a brief
overview of the Lagrangian formalism and standard Lagrangians.
Section 3 describes and discusses the models of the population 
dynamics, and Section 4 concludes the paper.

\section{Lagrangian formalism}

\subsection{General overview}

Preliminary formulation of the Lagrangian formalism was originally 
done by Euler in 1742, and then it was refined and applied to 
Newtonian dynamics by Lagrange, who set up its currently used 
form in his {\it Analytic Mechanics} that first appeared in 1788; 
however, see [3] for more recent edition.  According to Lagrange,
the formalism deals with a functional $\mathcal{S} [x(t)]$, which 
depends on a continuous and differentiable function $x(t)$ that 
describes evolution of a property of any dynamical system (given 
by $x$) in time (represented by $t$).  This evolution is described 
by an equation of motion, which can be derived from the Lagrangian
formalism. 

The functional $\mathcal{S}[x(t)]$ is called action and is defined by 
an integral over a scalar function $L$ that depends on $\dot x(t)= 
dx/dt$, $x$ and on $t$, so $L(\dot{x},x,t)$ and it is called the Lagrangian 
function or simply Lagrangian.  According to the principle of least action, 
or Hamilton's principle [4,5], the functional $\mathcal{S}[x(t)]$ must 
obey the following requirement $\delta\mathcal{S}=0$,  which says 
that the variation represented by $\delta$ must be zero to guarantee 
that the action is stationary (to have either a minimum, maximum 
or saddle point).  The necessary condition that $\delta\mathcal{S} 
=0$, is known as the Euler--Lagrange (E--L) equation that can be 
written as

\begin{equation}
{d\over{dt}}\left({{\partial L}\over{\partial{\dot x}}}\right) - 
{{\partial L}\over{\partial x}} = 0.
\label{eq1}
\end{equation}

The Euler--Lagrange equation leads to a second-order ordinary differential 
equation (ODE) that can be further solved to obtain $x(t)$ that makes the
action stationary.  The procedure forms the basis of calculus of variations, 
and it works well when the Lagrangian $L(\dot x,x,t)$ is already known.  
Deriving the second-order ODE from the E--L equation is called the {\it 
Lagrangian formalism}, and in this paper we deal exclusively with this 
formalism.  It must be pointed out that the formalism has been extensively 
used in modern physics and that all its fundamental equations are derived 
by using it [5].  In this paper, we develop the Lagrangian formalism in 
biology and use it to obtain equations of motions for the selected models 
of population dynamics. 

For dynamical systems whose total energy is conserved, the existence of 
Lagrangians is guaranteed by the Helmholtz conditions [6], which can also 
be used to obtain Lagrangians.  The procedure of finding Lagrangians is 
called the inverse (or Helmholtz) problem of calculus of variations [7] 
and it shows that there are three separate classes of Lagrangians, namely, 
standard [3,4], nonstandard [8] and null [9] Lagrangians.  Both standard 
and nonstandard Lagrangians give the same equations of motion after 
they are substituted into the E-L equation.  However, null Lagrangians 
satisfy the E-L equation identically and therefore they do not give any 
equation of motion.  

Our main goal is to establish the Lagrangian formalism for known ODEs 
that describe time evolution of different models of the population dynamics 
and find their Lagrangians; in this paper, we concentrate exclusively on 
standard Lagrangians.

\subsection{Standrard Lagrangians}

According to Lagrange [3], standard Lagrangians (SLs) are represented by 
differences between kinetic and potential energies of dynamical systems.
In dynamics, if an oscillatory system of mass $m$ is displaced by $x$ 
in time $t$, then its kinetic energy of oscillations is given by 

\begin{equation}
E_{\rm kin}={1\over 2}m\dot x(t),
\label{eq2}
\end{equation}

where $\dot x(t)$ is the time derivative of the dynamical variable $x(t)$, 
or otherwise known as $\dot x=dx/dt$.  The potential energy depends 
on a spring constant $k$ of the oscillatory system and it can be written as 

\begin{equation}
E_{\rm pot}=kx^2(t).
\label{eq3}
\end{equation}

Thus, the standard Lagrangian for this system is 

\begin{equation}
L(\dot{x},x)=E_{\rm kin}-E_{\rm pot}=
{1\over 2}\left[m\dot x(t)-kx^2(t)\right].
\label{eq4}
\end{equation}

Substitution of this Lagrangian into the E-L equation (see Eq. \ref{eq1}) 
gives the following equation of motion:

\begin{equation}
m\ddot x(t)+kx(t)=0,
\label{eq5}
\end{equation}

where $\ddot x(t)$ is the second derivative of $x(t)$ with respect to time,
or $\ddot x=d^2x/dt^2$.

The presented above form of the SL is characteristic for oscillatory systems 
that are not damped and not driven [2].  Obviously, the meaning of the 
variable $x(t)$ may change from one physical system to another.  Moreover,
the meaning of $x(t)$ in biological systems will be different than in physical 
systems, so will the meaning of the constants $m$ and $k$.  Nevertheless, 
a Lagrangian with a kinetic-like energy term, which contains $\dot x^2(t)$, 
and a potential-like energy term, which contains $x^2(t)$, will be identified 
as the SL.  The main objective of this paper is to derive SLs for different 
models of the population dynamics and discuss the role of these SLs in 
mathematical description of these models. 

The Lagrangian formalism based on standard Lagrangians have been 
well-established in most fields of modern physics (e.g., [5] and references
therein).  Specifically, the formalism is commonly used to obtain equations 
of motion for dynamical systems in Classical Mechanics [3,7,8].  More 
recently, several methods were developed to solve the inverse Helmholtz 
problems for physical systems described by ODEs (e.g., [10-14]).  Some 
of these methods developed for physical systems will be used in this 
paper to derive Lagrangians for biological systems. 

As already mentioned in the Introduction, there have been several attempts 
to establish the Lagrangian formalism in biology [15-18], and different 
Lagrangians were obtained for some selected biological systems.  In this 
paper, we concentrate on standard Lagrangians (SLs) because of their 
specific physical meaning discussed above.  Our goals are to derive SLs 
for some of the model for population dynamics, and discuss their meaning 
and role of these Lagrangians in biology.

\subsection{Method to derive standard Lagrangians}

The Lagrangian formalism requires prior knowledge of a Lagrangian.
In general, there are no first principle methods to obtain Lagrangians,
which are typically presented without explaining their origin.  In physics,
most dynamical equations were established first and only then their 
Lagrangians were found, often by guessing.  Once the Lagrangians 
are known, the process of finding the resulting dynamical equations 
is straightforward and it requires substitution of these Lagrangians 
into the E-L equation.   There has been some progress in deriving 
Lagrangians for physical systems described by ODEs (e.g., [10-14]), 
and in this paper, we develop a new method to obtain standard 
Lagrangians (SLs) for several models that describe interacting 
species of the population dynamics.  Let us point out that the 
same method can be used in physics and in other natural sciences 
to obtain Lagrangians for systems described by second-order ODEs. 

The main objective is to solve the inverse (Helmholtz) problem of 
the calculus of variations [6,7] and derive the standard Lagrangian
for a given second-order ODE.  Let us consider the following ODE 

\begin{equation}
\ddot x+\alpha (x)\dot x^2+\beta(x)\dot x+\gamma(x)x=C_0,
\label{eq6}
\end{equation}

where $\alpha(x)$, $\beta(x)$ and $\gamma(x)$ are at least twice 
differentiable functions of the dependent variable only, and these 
functions are to be specified by the selected models of the population 
dynamics.  In addition, $C_0$ is a constant that may appear in some
of these models (see Section 3).  From a physical point of view, the 
above equation of motion describes an oscillatory system that is 
affected by two damping terms with $\alpha (x)$ and $\beta (x)$,
and driven by the constant force $C_0$.  In case, $\alpha (x) = 
\beta (x) = C_0 = 0$, the equation represents a harmonic 
oscillator [4,5].  

Let us consider a simplified form of this ODE by taking $\beta (x)=0$
and $C_0 = 0$, and obtain

\begin{equation}
\ddot x+\alpha(x)\dot x^2+\gamma(x)x=0.
\label{eq7}
\end{equation}

Despite the presence of the damping-like term $\alpha (x)\dot x^2$,
the equation is considered to be conservative, which means that its 
standard Lagrangian can be obtained [11,20] and written as 

\begin{equation}
L(\dot x,x)={1\over 2}\dot x^2e{^{2I(x)}}-\int\displaylimits_{}^x
\tilde x\gamma(\tilde x)e^{2I(\tilde x)}d\tilde x,
\label{eq8}
\end{equation}

with 

\begin{equation}
I(x)=\int\displaylimits_{}^x\tilde\alpha(\tilde x)d\tilde x.
\label{eq9}
\end{equation}

The main reason for reducing Eq. (\ref{eq6}) to Eq. (\ref{eq7}) is 
that the term $\beta(x)\dot x$ is by itself a null Lagrangian
that identically satisfies the E-L equation [21-23], therefore, its 
derivation from any SL is not possible [11,12].  In the following, 
we propose to account for this term in a novel way.

We write Eq. (\ref{eq7}) in the following form

\begin{equation}
\ddot x+\alpha(x)\dot x^2+\gamma(x)x=F(x,\dot x),
\label{eq10}
\end{equation}

where the force-like term becomes

\begin{equation}
F(x,\dot x)=C_0-\beta(x)\dot x.
\label{eq11}
\end{equation}

All population dynamics models considered in this paper (see Section 3) 
are represented by second-order ODEs in the form of Eq. (\ref{eq10}) 
with $F(x,\dot x)$ given by Eq. (\ref{eq11}). 

Our results presented in the Appendix generalize the previous 
work [20] and demonstrate that the Lagrangian given by Eq. 
(\ref{eq8}) is also the SL for Eq. (\ref{eq10}) if, and only if, 
this SL is substituted into the E-L equation 

\begin{equation}
{d\over{dt}}\left({{\partial L}\over{\partial{\dot x}}}\right)-
{{\partial L}\over{\partial x}} = F(\dot x,x).
\label{eq12}
\end{equation}

This form of the E-L equation is commonly known in physics,
and the presence of the force-like term on the right-hand side 
of the equation is justified by $F(\dot x, x)$ being a force that 
does not arise from a potential (e.g., [4]); for further 
discussion of this term, see Section 3.3.   

It is straightforward to verify that substitution of Eq. (\ref{eq8})
into Eq. (\ref{eq12}) gives the original equation given by Eq. 
(\ref{eq10}).  This makes our method easy to find SLs for any 
ODEs of the form of Eq. (\ref{eq10}) as it requires identifying 
the functions $\alpha(x)$, $\gamma(x)$ and $F(\dot x,x)$, and 
evaluating the integral in Eq. (\ref{eq9}) as well as the one in 
the SL given by Eq. (\ref{eq8}).

Our method is simple and easy to use for the population dynamics 
models considered in this paper.  Let us point out that this method
can be applied to any dynamical system that is described by the 
second-order ODEs of the form of Eq. (\ref{eq10}) in any areas of 
natural sciences. 

\section{Applications to models of population dynamics}

\subsection{Selected models}

The models of the population dynamics considered in this paper are
listed in Table 1.  Our selection process was guided by the previous 
work, by Trubatch and Franco[17], and Nucci and Tamizhmani [18].  
Both papers considered the well-known population models that involve 
two interacting species described by coupled nonlinear ODEs, namely, 
the Lotka-Volterra, Gompertz, Verhulst and Host-Parasite models as 
shown in Table 1. The authors of these papers determined Lagrangians 
corresponding to the ODEs representing mathematically the models, either 
{\it ad hoc} [19] or using the method of Jacobi Last Multiplier [17]; the 
Lagrangians of the same form were obtained, and they were treated as 
the generating functions for the ODEs representing the models [17,18].  

As it is well-known, Lagrangians can be of different forms and yet 
they would give the same equation of motion [5,7].  In most cases, 
the forms of these Lagrangians do not resemble the SLs in which 
the kinetic and potential energy-like terms can be identified [10-14].
However, the main objective of this paper is to derive the SLs and    
for some selected models of the population dynamics and compare 
the obtained SLs to those previously found [18,19].  Because of the 
specific physical meaning of the SLs derived here, we are able to
address the role and meaning of these SLs in the population dynamics.

In selecting models of the population dynamics, we used the four models 
used in the previous studies [19,18]. In addition, we selected the SIR
model (see Table 1).

\renewcommand{\arraystretch}{1.3}
\begin{center}\begin{tabular}
{ll} \hline
{\bf Population models}	&{\bf Equations of Motion}\\ \hline
Lotka-Volterra Model &	$\dot{w_1}=w_1(a\:+\:bw_2\:)$\\
			&	$\dot{w_ 2}=w_2(A\:+Bw_1)$\\ \hline
Verhulst Model 	&	$\dot{w}_1=w_1(A+\:Bw_1\:+f_1 w_2)$\\
			&	$\dot{w}_2=w_2(\:a+\:bw_2\:+f_2 w_1)$\\ \hline
Gompertz Model 	&	$\dot{w}_1=w_1{(}A\log{\bigr(}{w_1\over m_1}
{\bigr)}+B{w_2}{)}$\\ 
			&	$\dot{w}_2=w_2{(}\:a\log{\bigr(}{w_2\over m_2}
{\bigr)}+b{w_1})$\\ \hline
Host-Parasite Model	&	$\dot{w}_1=w_1(a\:-bw_2)$\\ 
			&	$\dot{w}_2=w_2(A-B{w_2\over w_1})$ \\ \hline
SIR Model 		&	$\dot w_1=-b w_1w_2$\\
			&	$\dot w_2=\:\:\:b w_1w_2-aw_2$\\ \hline
\end{tabular}\end{center}

The first four models of the population dynamics presented in Table 1 describe 
two interacting species (preys and predators) of the respective populations 
$w_1 (t)$ and $w_2 (t)$ that evolve in time $t$, which is denoted by 
the time derivatives $\dot w_1 (t)$ and $\dot w_2 (t)$.  The coefficients 
$a$, $A$, $b$, $B$ $f_1$, $f_2$, $m_1$ and $m_2$ are real and constant
parameters that describe the interaction of the two species.  The Lotka-Volterra, 
Verhulst and Gompertz models are {\it symmetric}, which means that the 
depedent variables can be replaced if, and only if, the constants are replaced, 
$a \rightarrow A$, $b \rightarrow B$, $f_1 \rightarrow f_2$ and $m_1 
\rightarrow m_2$.  However, the Host-Parasite model is {\it asymmetric} 
in the dependent variables.  

The SIR model presented in Table 1 describes the spread of a disease in a 
population and the dependent variables $w_1(t)$ and $w_2(t)$ represent 
susceptible and infectious populations, with $a$ and $b$ being the recovery 
and infection rates, respectively.  Similarly to the Host-Parasite model, the 
SIR model is also {\it asymmetric} but the origin and nature of this 
asymmetry in both models is significantly different.

\subsection{Steps to derive standard Lagrangians}

We solve the inverse (Helmholtz) calculus variational problem and 
derive standard Lagrangians for all selected models of the population 
dynamics shown in Table 1.  The following steps must be undertaken 
to solve the problem:

\begin{enumerate}

\item Convert the set of coupled nonlinear ODEs for each model into a 
second-order nonlinear ODE for each dependent variable.

\item Cast the derived second-order ODEs into the equation of the same 
form as Eq. (\ref{eq10}).  

\item Compare the obtained second-order ODEs to Eq. (\ref{eq10}) 
and identify the functions $\alpha (x)$, $\gamma (x)$ and $F(x,\dot x)$ 
for each variable in each model.

\item Evaluate the integral in Eq. (\ref{eq9}) and then determine the 
exponential factor.

\item Evaluate the integral in Eq. (\ref{eq8}).

\item Use the above integrals and Eq. (\ref{eq8}) to find the standard
Lagrangian for each variable in each model.

\item Verify the derived standard Lagrangian by substituting it into the 
Euler-Lagrange equation given by Eq. (\ref{eq12}).

\end{enumerate}

As pointed out in [19], the choice between the first and second order 
description of the population dynamics is typically motivated by the 
information available.  In case only one population is observed, then
the second-order ODE can be solved for this population.  To make it
more general, we derive the second-order ODE for each variable in 
each model, so both equations can be solved if both populations are 
observed.  For this paper and the presented results, the second-order
ODEs are required because our method to derive the SLs is valid only 
for such equations.   
 
Our method to derive the SLs for the selected models is straightforward 
and the above steps are easy to perform and they always give the standard 
Lagrangian for any second-order ODEs of the same form as Eq. (\ref{eq10}).
In the following, we present the derived second-order ODEs for each variable 
of the model and the resulting SLs.

\subsection{Standard Lagrangians for selected models}

Our method to derive standard Lagrangians for the models presented in Table 1 
requires that the systems of coupled nonlinear first-order ODEs are cast into one 
second-order ODE for each variable.  All derived second-order ODEs can be 
expressed in the same form as Eq. (\ref{eq10}), which can be written as 

\begin{equation}
\ddot w_{i}+\alpha_{i}(w_{i})\dot w_{i}^2+\gamma_{i}(w_{i})w_{i}=
F_{i}(\dot w_{i}, w_{i}),
\label{eq13}
\end{equation}

where $i=1$ and $2$.   Since $w_i (t)$ represents the population of species,
its derivative with respect of time $\dot w_i (t)$ describes the rate with which 
the population changes, and $\ddot w_i (t)$ its acceleration.  Despite the 
presence of the damping-like term $\alpha_{i}(w_{i})\dot w_{i}^2$, the 
LHS of the above equation is conservative [11,20] and it describes oscillations 
of the population of species with respect to its equilibrium.  These oscillations
are modified by the force-like term on the RHS of the equation.  Let us now 
describe this term.

Typically, the presence of any term with $\dot w_i (t)$ corresponds to friction 
forces in classical mechanics.  In the approach presented in this paper, all 
friction-like terms that explicitly depend on $\dot w_i (t)$ are collected on the 
RHS of the equation as $F_{i} (\dot w_{i}, w_{i})$, which becomes the force-like 
term.  Since $F_{i} (\dot w_{i}, w_{i})$ arises directly from the friction-like 
terms, its origin is not potential, and therefore this force-like term may appear 
on the RHS of the E-L equation (see Eq. \ref{eq12}) as it is shown in [4].

In our derivations of the standard Lagrangians for the models of the population 
dynamics presented in Table 1, we follow the above steps.  As an example, we 
show the calculations required for each step for the Lotka-Volterra model.  
However, for the remaining four models in Table 1, we only present the 
final results. 

\section{Lotka-Volterra Model}

The Lotka-Volterra model was developed by chemist Alfred Lotka in 1910 and 
mathematician Vito Volterra in 1926 [24,25], and this model describes the interaction
of two populations (predator-prey) based on the assumptions that the prey increases
exponentially in time without the predator, and the predator decreases exponentially 
without the prey [19].  The model is symmetric and it is represented mathematically 
by a system of coupled nonlinear ODEs given in Table 1.

{\it Step 1:} Using the coupled nonlinear ODEs for this model, we convert the equations 
into the following second-order ODEs for the variables $w_1$ and $w_2$

\begin{subequations}        
\begin{equation}
\ddot w_1-{\dot w_1^2\over w_1}-B\dot{w_1}w_1-A\dot{w_1}
+aBw_1^2 + aA{w_1}=0,
\label{eq13a}
\end{equation}

and

\begin{equation}
\ddot w_2-{\dot w_2^2\over w_2}-bw_2\dot w_2-a\dot{w_2}+Abw_2^2+
Aa{w_2}=0.
\label{eq13b}
\end{equation}
 \end{subequations}

{\it Step 2:} We cast these equations into the form of Eq. (\ref{eq13}), and obtain 

\begin{subequations}        
 \begin{equation}
\ddot w_1\underbrace{-{1\over w_1}}_{\alpha_1(w_1)}\dot w_1^2+
\underbrace{(Bw_1+A)a}_{\gamma_1(w_1)}w_1=\underbrace{(Bw_1+A)
\dot w_1}_{F_1(\dot w_1,w_1)}
\label{eq14a}
\end{equation}  

and

\begin{equation}
\ddot w_2\underbrace{-{1\over w_2}}_{\alpha_2(w_2)}\dot w_2^2+
\underbrace{(bw_2+a)A}_{\gamma_2(w_2)}w_2=\underbrace{(bw_2+a)
\dot w_2}_{F_2(\dot w_2,w_2)}
\label{eq14b}
\end{equation}
\end{subequations}

{\it Step 3:} Using the above equations, we identify the functions 

\renewcommand{\arraystretch}{1.4}
\begin{center}
\begin{tabular}{l  l}
$\alpha_1(w_1)=-{1\over w_1}$\\
$\alpha_2(w_2)=-{1\over w_2}$\\
$\gamma_1(w_1)=a\:(Bw_1+A)$\\
$\gamma_2(w_2)=A\:(bw_2+a)$\\
$F_1(\dot w_1,w_1)=(Bw_1+A)\dot w_1$\\	
$F_2(\dot w_2,w_2)=(\:bw_2+a)\dot w_2$\\
\end{tabular}
\end{center}

{\it Step 4:} Having identified $\alpha(w_1)$ and $\alpha(w_2)$, the intergral 
given by Eq. (\ref{eq9}) can be evaluated, and the result is  

\begin{subequations}
\begin{equation}
I(w_1)=-{\int\displaylimits^{\tilde w_1}}{dw_1\over w_1}=-\ln|w_1|,	
\label{eq15a}
\end{equation}

and 

\begin{equation}
I(w_2)=-{\int\displaylimits^{\tilde w_2}}{dw_2\over w_2}=-\ln|w_2|.
\label{eq15b}
\end{equation}
\end{subequations}

Then, the factors $e^{2I(w_1)}$ and $e^{2I(w_2)}$ become

\begin{subequations}
\begin{equation}
e^{2I(w_1)}=e^{-2\ln|w_1|}={1\over w_1^2},
\label{eq16a}
\end{equation}

and

\begin{equation}
e^{2I(w_2)}=e^{-2\ln|w_2|}={1\over w_2^2}.
\label{eq16b}
\end{equation}
\end{subequations}

{\it Step 5:} Since $\gamma(w_1)$ and $\gamma(w_2)$ are known, the 
intergral in Eq. (\ref{eq8}) can be calculated, and we find

\begin{subequations}
\begin{equation}
\int\displaylimits_{}^{w_1}\tilde{w_1}\gamma(\tilde {w_1})e^{2I(\tilde {w_1})}
d\tilde {w_1}=a(Bw_1+A\ln|w_1|),
\label{eq17a}
\end{equation}

and

\begin{equation}
\int\displaylimits_{}^{w_2}\tilde{w_2}\gamma(\tilde {w_2})e^{2I(\tilde {w_2})}
d\tilde{w_2}=A(bw_2+a\ln|w_2|).
\label{eq17b}
\end{equation}
\end{subequations}

{\it Step 6:} Using the above results and the definition of Lagrangian 
given by Eq. (\ref{eq8}), the following standard Lagrangians for the 
two dependent variables of the Lotka-Volterra model are obtained: 

\begin{subequations}
\begin{equation}
L_1(\dot w_1,w_1)=\underbrace{{1\over 2}\biggl({\dot w_1\over w_1}
\biggl)^2}_{kinetic\ \rm term}-\underbrace{a\bigr(Bw_1+A
\ln|w_1|\bigr)}_{potential\ \rm term}
\label{eq18a}
\end{equation}

and

\begin{equation}
L_2(\dot w_2,w_2)=\underbrace{{1\over 2}\biggl({\dot w_2\over w_2}
\biggl)^2}_{kinetic\ \rm term}-\underbrace{A\bigr(bw_2+a
\ln|w_2|\bigr)}_{potential\ \rm term}
\label{eq18b}
\end{equation}
\end{subequations}

{\it Step 7:} Substituting the derived standard Lagrangians and $F(\dot w_i, 
w_i)$ into the following Euler-Lagrange equations 

\begin{equation}
{d\over{dt}}\left({{\partial L}\over{\partial {\dot w_i}}}\right) - 
{{\partial L}\over{\partial w_i}} = F(\dot w_i,w_i)e^{2I(w_i)},
\label{eq19}
\end{equation}

where $i=1$ and $2$, we obtain Eqs. (\ref{eq14a}) and (\ref{eq14b}).  
This verifies that the presented method to derive the SLs is valid.

\section{Verhulst Model}

This logistical (or Verhulst-Pearl) equation was first introduced by the Belgian 
statistician Pierre Francois Verhulst in 1838 [27].  This model describes the 
organisms' growth dynamics in a habitat of finite resources, which means the 
population is limited by a carrying capacity. This model is valueable for optimisation
of culture media by developing strategies and selection of cell lines.

In this paper, the Verhulst model describes the population of interacting species 
by considering self-interacting terms that prevent the exponential increase or 
decrease in the size in the populations observed in the Lotka-Volterra model [19].
The system of coupled nonlinear ODEs given in Table 1 shows that this model is 
symmetric.

After performing steps 1 through 3, the following functions for this model are obtained: 

\begin{center}
\begin{tabular}{l l}
$\alpha_1(w_1)=-(1+\:b\:){1\over w_1}$\\ 
$\alpha_2(w_2)=-(1+B){1\over w_2}$\\
$\gamma_1(w_1)=\bigr(f_2-b\:\bigr)Bw_1^2+\bigr(Af_2-2Ab-a)w_1+A\bigr(a-Ab\bigr)$\\
$\gamma_2(w_2)=\bigr(f_1-B\bigr)Bw_2^2+\bigr(\:af_1-2ab-A)w_1+a\bigr(A-aB\bigr)$\\
$F_1(\dot w_1,w_1)=-\dot w_1\bigr[(2b-1)Bw_1-f_2w_1^2+(2Ab-a)+(f_2-b)B\bigr]$\\
$F_2(\dot w_2,w_2)=-\dot w_2\bigr[(2B-1)bw_2-f_1w_2^2+(2aB-A)+(f_1-B)b\bigr]$\\
\end{tabular}
\end{center}

Substitution of these functions into Eq. (\ref{eq13}) gives the second-order ODEs for the 
variables $w_1$ and $w_2$.

Then, we implement steps 4 through 6 and the resulting standard Lagrangians for the 
variables $w_1$ and $w_2$ of the Verhulst model are 

\begin{subequations}
\begin{equation}
\resizebox{.8\hsize}{!}{$
L(\dot w_1,w_1)={1\over 2}\biggl[\biggl({\dot w_1\over w_1}\biggl)^2-{\:(f_2-b)B\over(1-b)}
w_1^2-{2(Af_2-2Ab-a)\over(1-2b)}w_1+{A(a-Ab)\over b}\biggl]w_1^{-2b}$}
\label{eq20a}
\end{equation}

and

\begin{equation}
\resizebox{.8\hsize}{!}{$
L(\dot w_2,w_2)={1\over 2}\biggl[\biggl({\dot w_2\over w_2}\biggl)^2-{(f_1-B)b\over(1-B)}
w_2^2-{2(af_1-2aB-A)\over(1-2B)}w_2+{a(A-aB)\over B}\biggl]w_2^{-2B}$}
\label{eq20b}
\end{equation}
\end{subequations}

The kinetic and potential energy-like terms are easy to recognize, and the functions
$F_1(\dot w_1,w_1)$ and $F_2(\dot w_2,w_2)$ are given above.  Substitution of 
these Lagrangians into the E-L equations (see Eq. \ref{eq19}) validates the method.

\section{Gompertz Model}

English economist Benjamin Gompertz proposed a model to describe the relationship 
between increasing death rate and age in 1825. This model is useful in describing 
the rapid growth of a certain population of organisms such as the growth of tumors [28].
As well as, modelling the amount of medicine in the bloodstream.  

Here, we follow [17,19] and consider the Gompertz model for the population dynamics
This model generalizes the Lotka-Volterra model by including self-interaction terms that 
prevent an unbounded increase of any isolated population [19]; the self-interacting terms 
in the Gompertz model are different than those in the Verhulst model.  The mathematical 
representation of this model given by the coupled and nonlinear ODEs in Table 1 shows 
that the model is symmetric.

The steps 1 through 3 allow us to identify the following functions in Eq. (\ref{eq13})

\begin{center}
\begin{tabular}{l}
$\alpha_1(w_1)=-{1\over w_1}$\\
$\alpha_2(w_2)=-{1\over w_2}$\\			
$\gamma_1(w_1)=\bigr[A\log\bigr({w_1\over m_1}\bigr)\bigr]w_1$\\
$\gamma_2(w_2)=\bigr[a\:\log\bigr({w_2\over m_2}\bigr)\bigr]w_2$\\
$F_1(\dot w_1,w_1)=[Am_1+\:bw_1+g_1(\dot w_1,w_1)] \dot w_1-g_1(\dot w_1,w_1)Aw_1$\\
$F_2(\dot w_2,w_2)=[\:am_2+Bw_2+g_2(\dot w_2,w_2) ]\dot w_2-g_2(\dot w_2,w_2)aw_2$\\
\end{tabular}
\end{center}

where 

\begin{subequations}
\begin{equation}
g_1(\dot w_1,w_1)\:=\:a\log\:\biggr[{1\over m_2B}\:\biggr({\dot w_1\over w_1}-
A\log\biggr({w_1\over m_1}\biggr)\biggr)\biggr],
\end{equation}
and 

\begin{equation}
g_2(\dot w_2,w_2)\:=A\log\:\biggr[{1\over m_1b}\:\:\biggr({\dot w_2\over w_2}-
a\log\biggr({w_2\over m_2}\biggr)\biggr)\biggr].
\end{equation}
\end{subequations}

Then, the remaining steps result in the following standard Lagrangians for the two 
dependent variables of the Gompertz model

\begin{subequations}
\begin{equation}
L_1(\dot w_1,w_1)={1\over 2}\biggr({\dot w_1\over w_1}\biggr)^2-A\biggr[\log
\biggr({w_1\over m_1}\biggr)-1\biggr]w_1
\label{eq21a}
\end{equation}

and

\begin{equation}
L_2(\dot w_2,w_2)={1\over 2}\biggr({\dot w_2\over w_2}\biggr)^2-a\biggr[\log
\biggr( {w_2\over m_2}\biggr)\:-1\biggr]w_2.
\label{eq21b}
\end{equation}
\end{subequations}

In both Lagrangians the kinetic and potential energy-like terms are seen, and the 
forcing functions $F_1(\dot w_1,w_1)$ and $F_2(\dot w_2,w_2)$ are given above.
If we substitute these Lagrangians into Eq. (\ref{eq19}), the second-order ODEs 
for the variables $w_1$ and $w_2$ are obtained.

\section{Host-Parasite Model}

This model describes the interaction between a host and its parasite.
The model takes into account nonlinear effects of the host population 
size on the growth rate of the parasite population [19].  The system 
of coupled nonlinear ODEs (see Table 1) is asymmetric in the dependent
variables $w_1$ and $w_2$.

Using steps 1 through 3, we find 

\begin{center}
\begin{tabular}{l l}
$\alpha_1(w_1)=-{1\over w_1}\Bigr(1+{B\over bw_1}\Bigr)$\\ 
$\alpha_2(w_2)=-{2\over w_2}$\\
$\gamma_1(w_1)=aA$\\
$\gamma_2(w_2)=A(bw_2-a)$\\
$F_1(\dot w_1,w_1)=B{\:\:a^2\over b}\:+\Bigr(A-{2aB\over bw_1}\Bigr)\dot w_1$\\	
$F_2(\dot w_2,w_2)=(bw_2-a-A)\dot w_2$
\end{tabular}
\end{center}

After substituting these functions into Eq. (\ref{eq13}), the second-order ODEs for the 
variables $w_1$ and $w_2$ are obtained, and the equations are asymmetric, which means
that the remaining steps 4 through 6 must be applied to each dependent variable ($w_1$
or $w_2$) separately.  

The standard Lagrangian for the variable $w_1$ is 

\begin{subequations}
\begin{equation}
L_1(\dot w_1,w_1)={1\over 2}\Biggr({\dot w_1\over w_1}\Biggr)^2e^{2B/bw_1}+
aAEi\:\Biggr({2B\over bw_1}\Biggr),
\label{eq22a}
\end{equation}
\end{subequations}

where the exponential integral $Ei(2B/bw_1)$ is a special function defined as 

\begin{equation}
Ei(z) = \int_{\infty}^{z}{{e^{\tilde z}} \over {\tilde z}} d\tilde z,
\end{equation}

with $z=2B/bw_1$.  It must be noted that $Ei(z)$ is not an elementary 
function.  Now, the standard Lagrangian for the variable $w_2$ is given by 

\begin{subequations}
\begin{equation}
L_2(\dot w_2,w_2)={1\over 2}\Biggr({\dot w_2\over w_2}\Biggr)^2{1\over w_2^2}-
A\Biggr[{1\over 2}{a\over w_2}-b\Biggr]{1\over w_2}.
\label{22b}
\end{equation}
\end{subequations}

It is seen that there are significant differences between the Lagrangian for $w_2$ and 
that for $w_1$ in both the kinetic and potential energy-like terms.  The differences are 
especially prominent in the potential energy-like terms, whose explicit dependence on 
the exponential integral $Ei(2B/bw_1)$ is a new phenomenon.  The differences are 
caused by the asymmetry between the dependent variables in the original equations 
(see Table 1), which makes this model different than the fully symmetric Lotka-Volterra, 
Verhulst and Gompertz models, whose standard Lagrangians are also fully symmetric.

\begin{subequations}
\begin{equation}
g_1(\dot w_1,w_1)\:=\:a\log\:\biggr[{1\over m_2B}\:\biggr({\dot w_1\over w_1}-
A\log\biggr({w_1\over m_1}\biggr)\biggr)\biggr],
\label{eq27a}
\end{equation}

and

\begin{equation}
g_2(\dot w_2,w_2)\:=A\log\:\biggr[{1\over m_1b}\:\:\biggr({\dot w_2\over w_2}-
a\log\biggr({w_2\over m_2}\biggr)\biggr)\biggr].
\label{eq27b}
\end{equation}
\end{subequations}

\section{SIR Model}

Kermack and McKendrick in 1927 derived the system of the first ODEs (see Table 1) 
describing the spread of a disease in a population [29]. It is the one of the simplest 
model, dividing the population into three distinct sub-populations: a susceptible 
population denoted by $w_1(t)$, the infectious population represented by $w_2(t)$, 
and a recovered population, we denote as $w_3(t)$.  

It is seen that the dependent variable $w_3(t)$ does not appear explicitly in the set 
of ODEs given in Table 1 because it is related to $w_1(t)$ and $w_2(t)$ through 
the following population conservation law:  $d/dt(w_1 + w_2 + w_3)=0$, which 
means that the sum of the three populations must remain constant in time.  Moreover, 
$a>0$ is the recovery rate and $b>0$ is the rate of infection, which means that the 
terms $-bw_1w_2$ and $-aw_2$ represent newly infected and recovered individuals,
respectively.

After performing steps 1 through 3, we obtain 

\begin{center}
\begin{tabular}{l l}
$\alpha_1(w_1)= -{1\over w_1}$\\ 
$\alpha_2(w_2)= -{1\over w_2}$\\
$\gamma_1(w_1)=0$\\
$\gamma_2(w_2)=abw_2$\\
$F_1(\dot w_1,w_1)=(bw_1 - a)\dot w_1$\\	
$F_2(\dot w_2,w_2)=-bw_2\dot w_2$\\
\end{tabular}
\end{center}

Then, steps 4 through 6 give the following standard Lagrangians 

\begin{subequations}
\begin{equation}
L_1(\dot w_1,w_1)={1\over 2}\biggr({\dot w_1\over w_1}\biggr)^2,
\label{a}
\end{equation}

and

\begin{equation}
L_2(\dot w_2,w_2)={1\over 2}\biggr({\dot w_2\over w_2}\biggr)^2-abw_2.
\label{b}
\end{equation}
\end{subequations}

The fact that the SIR model is asymmetric is shown the lack of the potential
energy-like term in $L_1(\dot w_1,w_1)$ and its presence in $L_2(\dot w_2,
w_2)$.  However, the kinetic energy-like terms are the same for the SLs for 
both variables, and they are also similar to such terms in the SLs obtained 
for the other population dynamics models.

\subsection{Comparisons of Lagrangians and models} 

The derived standard Lagrangians for the considered models of the population 
dynamics are characterized by the terms corresponding to the kinetic and potential
energy as well as to the forcing function.   These terms are easy to identify (see 
Eqs. \ref{eq18a} and \ref{eq18b}) and they can be used to make comparisons  
between the Lagrangians and models they represent.   The models considered 
in this paper can be divided into two families, namely, symmetric (Lotka-Volterra, 
Verhulst and Gompertz) and asymmetric (Host-Parasite and SIR) models.  The 
SLs derived for these models are different than the Lagrangians previously 
obtained [19,17]; the main difference is the explicit time-dependence of those 
Lagrangians as compared to the SLs derived in this paper.  In the following, 
we describe the general characteristics of the derived SLs and comment on 
their explicit time-independence.

The kinetic energy-like terms in all four models have the same factor $(\dot w_i / 
w_i)^2 / 2$, where $i$ = 1 and 2, which represents the ratio at which the 
population changes to its value at a given time.  For the Verhulst and Host-Parasite 
models, this ratio is modified by the other factors that depend on the concentration 
of species at a given time.  It is interesting that the kine tic energy-like terms in the 
Lotka-Volterra, Gompertz and SIR models are independent from any constant 
parameters but for the other two models they are; in case of the Host-Parasite 
models only the variable $w_1$ shows such a dependence.

The potential energy-like terms of the Lotka-Volterra model depends linearly 
on the concentration of species; however, the Verhulst, Gompertz and 
Host-Parasite models also have nonlinear (second-order) terms in the 
concentration of species.  The SIR model is exceptional as its SL for the 
variable $w_1$ does not depend on any potential energy-like term.  On the 
other hand, the SL for the variable $w_2$ does depend on the potential 
energy-like term that is linear in this variable.  In all models, the potential 
energy-like terms depend on the constant parameters that appear in the 
derived second-order ODEs for these models.  An interesting result is the 
presence of logarithmic terms in the Lotka-Volterra and Gompertz models, 
and the exponential integral $Ei$ for the variable $w_1$ for the Host-Parasite 
model.  It must be also noted that the form of the potential energy-like term 
for the SIR model is the simplest among all the models considered here.
 
Since the kinetic and potential energy-like terms depend on the square of the 
rate with which the populations change and the square of their concentration,
respectively, one may conclude that the derived SLs describe oscillatory systems.
The fact that none of the obtained SLs depends explicitly on time indicates that 
the considered models oscillate in time with certain frequencies around the 
equilibrium, and that these oscillations are not directly affected by any 'physical 
damping' associated with the presence of terms that depend on $\dot w_i$, 
which are absent in derived the SLs; this is the main difference between the 
SLs of this paper and those previously obtained [19,17].  

However, we must keep in mind that all damping terms represented by $\dot w_i 
(t)$ have been moved to the force-like functions denoted by $F_i (\dot w_i, w_i)$, 
which significantly vary for different models.  First, let us point out that the 
force-like functions become null Lagrangians [21-23], which means that they 
identically satisfy the E-L equation, and therefore they cannot contribute to any 
equation of motion.  As a result, no standard Lagrangian can properly account 
for them [11,12] because the presented Lagrangian formalism is valid only for 
conservative systems.  

Second, the force-like functions may depend only on $\dot w_i (t)$ and on 
$\dot w_i (t) w_i (t)$, and the constant parameters, or may depend on higher
powers of thess variables that are arguments of the logarithmic functions.  
As the presented results demonstrate, the forms of the force-like functions 
significantly differ for different models, with the simplest being for the SIR 
and Lotka-Volterra models, and then the increasing complexity for the 
Host-Parasite and Verhulst models.  The most complex form of 
$F_i (\dot w_i, w_i)$ is found for the Gompertz model.  

Third, the observed increase in complexity of the force-like function is 
caused by the role played by the terms that depend on $\dot w_i$ as 
well as on the combination of terms with $\dot w_i w_i$.  In other 
words, the forms of $F_i (\dot w_i, w_i)$ significantly affect the 
oscillatory behavior of the systems represented by the SLs (see 
discussion above).   The force-like function may modify this 
behavior by causing the systems to reach the equilibrium faster 
or diverge from it; detailed analysis requires solutions to the ODEs
representing the considered systems, which is out of the scope of 
this paper.

Finally, let us point out that since the energy-like terms in the SLs
have specific meanings for the models, we suggest that they may
be used to identify physical similarities between different models 
as well as they may be utilized to classify the models into categories
that have similar biological characteristics.  Despite the fact that the 
SLs were derived only for the population dynamics models, the 
developed method and the above discussion can be easily applied 
to a broad range of biological systems, which will be explored in 
succeeding papers.
    
\section{Conclusions}

We developed Lagrangian formalism for the following population 
dynamics models: Lotka-Volterra, Verhulst, Gompertz, Host-Parasite 
and SIR models.  For ODEs that represent these models, we solved
the inverse (Helmholtz) variational calculus problem and derived 
standard Lagrangians for the models.  The main characteristic
of these Lagrangians is that their kinetic and potential energy-like
terms are identified and that they can be used to make comparisons
between the obtained Lagrangians as well as the models.  

The comparison demonstrates the role of these terms in the population
dynamics models and gives new insights into the models by showing 
their similarities and differences.  Moreover, the analogy between 
the derived standard Lagrangians and that known for a harmonic 
oscillator in physics is used to discuss the oscillatory behavior of 
the models with respect to their equilibrium.   In our approach, 
we collected the terms with the first-order derivatives in time and 
identified them as the force-like functions, which happened to be 
significantly different for each model and strongly depend on the 
parameters of each model.  By separating the force-like functions, 
we were able to see the effects of these functions on the model's 
oscillatory behavior.

Our method of solving the inverse calculus of variation problem
and deriving standard Lagrangians is applied to the models of 
population dynamics.  However, the presented results show 
that the method can be easily extended to other biological 
systems whose equations of motion are known.

\end{document}